\documentstyle[12pt,a4,epsf]{article}

\newcommand{\news}{\setcounter{equation}{0}}

\def\eqn{\begin{equation}}
\def\eeqn{\end{equation}}
\def\arr{\begin{array}}
\def\earr{\end{array}}
\def\eqna{\begin{eqnarray}}
\def\eeqna{\end{eqnarray}}
\def\a{\alpha}
\def\b{\beta}
\def\ha{\hat{\alpha}}
\def\hb{\hat{\beta}}
\def\s{\sigma}
\def\d{\delta}
\def\w{\wedge}
\def\o{\omega}
\def\e{\epsilon}

\def\z{\zeta}
\def\t{\tau}
\def\p{\partial}
\def\g{\gamma}
\font\mybb=msbm10 at 12pt
\def\bb#1{\hbox{\mybb#1}}

\def\bR {\bb{R}}
\def\bE {\bb{E}}

\def\ov {\overline}

\begin{document}

\vspace*{-.6in}
\thispagestyle{empty}
\begin{flushright}
DAMTP-R/97/59\\
\end{flushright}

{\Large
\begin{center}
{\bf Landau degeneracy and
black hole entropy}
\end{center}}
\vspace{.3in}
\begin{center}
Miguel S. Costa\footnote{M.S.Costa@damtp.cam.ac.uk}
and Malcolm J. Perry\footnote{malcolm@damtp.cam.ac.uk}\\
\vspace{.1in}
\emph{D.A.M.T.P.\\ University of Cambridge \\ Cambridge CB3 9EW \\ UK}
\end{center}

\vspace{.5in}

\begin{abstract}
We consider the supergravity solution describing a configuration of
intersecting D-4-branes with non-vanishing worldvolume gauge fields.
The entropy of such a black hole is calculated in terms of the
D-branes quantised charges. The non-extreme solution is also
considered and the corresponding thermodynamical quantities are
calculated in terms of a D-brane/anti-D-brane system. To perform the
quantum mechanical D-brane analysis we study open strings with their
ends on branes with a magnetic condensate. Applying the results
to our D-brane system we managed to have a perfect agreement between
the D-brane entropy counting and the corresponding semi-classical
result. The Landau degeneracy of the open string states
describing the excitations of the D-brane system enters in a crucial
way. We also derive the near-extreme results which agree with 
the semi-classical calculations.
\end{abstract}

\newpage

\section{Introduction}
\news

It is now widely accepted that superstring theory is the most
promising unifying theory. Among its great achievements is the
resolution of a longstanding problem in black hole physics: the
statistical origin of the Bekenstein-Hawking entropy at least for
certain extreme and near-extreme black holes
\cite{StroVafa,CallMald,Brec..1,MaldStro,John..,Mald}. The
resolution of  this problem has been made possible by the realization
of Polchinski \cite{Polc,Polc1} that D-branes, i.e. branes carrying
Ramond-Ramond charge, are extended objects with the property that open
strings can end on them. The D-brane excitations are described by such
open strings. The microscopic origin of the black hole entropy arises
in D-brane physics in two steps: firstly, we identify a certain
classical supersymmetric black hole carrying charge in the $R\otimes R$
sector with a corresponding excited D-brane configuration; secondly,
we count the number of open string states that reproduce the excited
D-brane system.

The worldvolume bosonic field content that describes the low energy
excitations of a D-$p$-brane is given by $9-p$ scalar fields $\phi^m$ 
($m=p+1,...,9$) and by a 1-form gauge potential $A_{\a}$
($\a=0,...,p$) \cite{Dai..,Leig}. These fields are interchanged by
$T$-duality transformations. The scalar fields are associated to the
massless excitations of the brane that describe transverse displacements,
and the gauge field is associated to worldvolume massless excitations
on the brane. Through this paper we will be interested in non-threshold
D-brane bound states. These configurations arise whenever the
worldvolume gauge field associated with a D-$p$-brane forms a
condensate. If this is the case the D-$p$-brane may carry the $R\otimes
R$ charge that is usually associated with lower dimensional
D-branes. In fact, the D-$p$-brane effective action contains
the correct coupling terms to $R\otimes R$ gauge potentials with rank
lower than $p+1$ \cite{Doug}. Remarkably, the existence of these
generalised Wess-Zumino terms (and for the D-9-brane the anomalous
Bianchi identity and the G-S coupling) is a consequence of $T$-duality
and Lorentz invariance \cite{Doug,Bach}.

To be definite consider the $(p|p-2)$ D-brane bound state. This
configuration may be obtained by performing a $T$-duality at an angle on
a D-$(p-1)$-brane, i.e.
\eqn
\arr{ccc}
\arr{rl}
p-1:& X^1,...,X^{p-1}\\
    & \phi^p =\tan{\z}\ X^{p-1}
\earr
&\longrightarrow &
\arr{rl}
p:& X^1,...,X^{p}\\
    & A_p =\tan{\z}\ X^{p-1}
\earr
\earr
\label{1.1}
\eeqn
where we start with a D-$(p-1)$-brane parallel to the $x^1, ...,
x^{p-2}$-directions and making an angle $\z$ with the
$x^{p-1}$-direction in the $x^{p-1}x^p$ 2-torus. The
$T$-duality transformation
along the $x^p$-direction interchanges the scalar field $\phi^p$ with
the gauge field $A_p$ forming a magnetic condensate (through this
paper we shall scale the gauge potential as $A_{\a}\equiv
2\pi\a'A_{\a}$). From the
supergravity theory perspective the existence of these bound states
has been deduce in \cite{Brec..,CostaPapa}.
 
In the linear approximation the field theory describing the low-lying
states on the D-brane bound state in (\ref{1.1}) is the super Yang
Mills theory with 
$U(m)$ gauge group \cite{Witt1}. The integer $m$ is the D-$p$-brane
winding number in the $x^{p-1}$-direction. The brane
carries a $U(1)$ flux which 
induces a t'Hooft twist in the $SU(m)/Z_m$ part of the theory
\cite{tHoo,tHoo1,GuraRamg,HashTayl}. As a consequence the gauge potential
and the scalar fields will obey twisted boundary conditions. For an
appropriate choice of this $U(m)$ bundle multiple transition functions
the only non-vanishing gauge potential is in the $U(1)$ center of the
group and it is given by (\ref{1.1}).
Had we considered $N$
D-branes on top of each other, the scalar and gauge fields would become
$U(Nm)$ matrices and equation (\ref{1.1}) would still refer to the
$U(1)$ center of the group.

The aim of this paper is to study the D-brane entropy counting when
the D-brane system has non-vanishing worldvolume gauge fields. In
particular, we will be considering a D-brane configuration presented
in \cite{CostaCvet} interpreted as the intersection of $N_1$ and $N_2$
D-$4$-branes with non-vanishing worldvolume gauge fields and carrying
momentum along a common direction. Although this configuration is
$T$-dual to the configuration used by Callan and Maldacena
\cite{CallMald}, the origin of the statistical
entropy has a remarkable new  feature. Namely, we have to take into
account the Landau degeneracy of the open string sector corresponding
to open strings with each end on different branes.

We begin in section two by presenting our D-brane configuration and by
considering the corresponding charges quantisation. We shall then
present the supergravity solution describing our D-brane system and
write the integer valued form of the Bekenstein-Hawking
entropy. The supergravity solution corresponding to the non-extreme
configuration will also be presented and the corresponding thermodynamical
formulae in terms of a brane/anti-brane system will be written. In
section three we shall consider open strings with ends on different
D-branes which have a magnetic condensate on its worldvolume theory. We
will keep our discussion general and specify for our D-brane system in
section four, where we shall give the D-brane entropy counting. We
will also consider the near-extreme black hole solution, matching the
corresponding thermodynamical quantities with the brane/anti-brane
system. We shall give our conclusions in section five.

\section{D-brane system}
\news

\begin{figure}[t]
\epsfxsize=5in
\centerline{\epsfbox{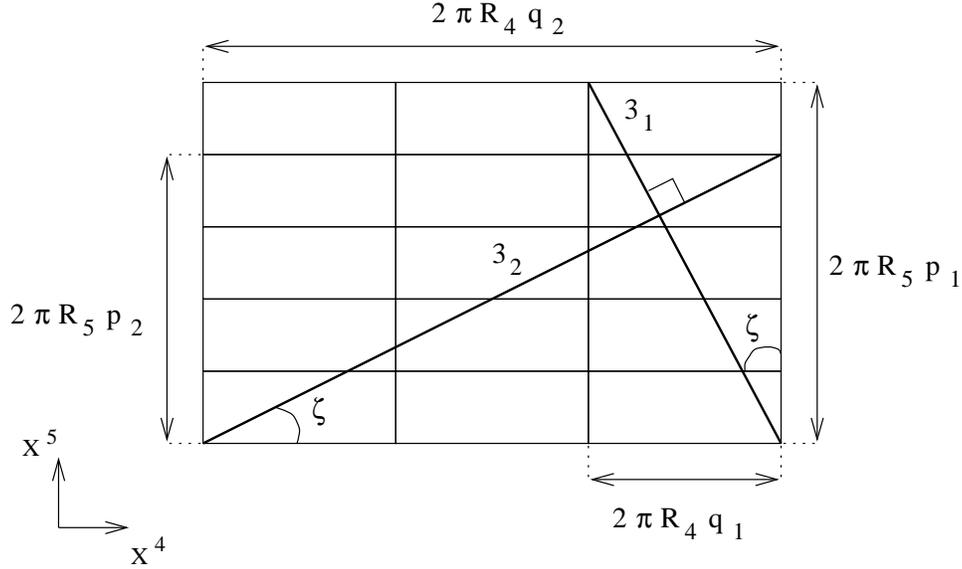}}
\caption{{\footnotesize $3 \perp 3$ D-brane system with each group of
    D-$3$-branes wrapped on the $(-q_1,p_1)$- and $(q_2,p_2)$-cycles of
  the $x^4x^5$ $2$-torus. In this example $q_1=1,\ p_1=5$ and $q_2=3$,
  $p_2=4$. The moduli $R_4$ and $R_5$ will then be related by
  $(R_5/R_4)^2=3/20$.}} 
\label{torus}
\end{figure}

Let us start by considering the $3\perp 3$ D-brane configuration. We
place $N_1$ D-$3$-branes along the $x^1, x^2, x^5$-directions and
$N_2$ D-$3$-branes along the $x^1, x^3, x^4$-directions. These
directions are taken to be circles of radius $R_i$. Next, we rotate
this configuration in the $x^4x^5$ $2$-torus by an angle $\z$
as described in figure \ref{torus}. The resulting D-brane system may be
represented by
\eqn
\arr{rl}
3_1:&X^1,X^2,X^5,\\
    &\phi^4_{(1)}=-\tan{\z}\ X^5\ ,\\
3_2:&X^1,X^3,X^5,\\
    &\phi^4_{(2)}=\cot{\z}\ X^5\ ,
\earr
\label{2.1}
\eeqn
where as explained in the introduction we are considering the $U(1)$
center of the worldvolume scalar fields associated with each group of
D-$3$-branes. Wrapping a direction of the D-$3$-branes around a cycle
of the $x^4x^5$ $2$-torus yields
quantisation conditions on the rotation angle $\z$. Start by
considering the first D-$3$-branes. If the
corresponding winding number in the $x^4$-direction is $-q_1$, and in
the $x^5$-direction is $p_1$, the angle $\z$ will obey the quantisation
condition
\eqn
\tan{\z}=\frac{q_1}{p_1}\frac{R_4}{R_5}\ ,
\label{2.2}
\eeqn
where $q_1$ and $p_1$ are co-prime. Similarly, if the other
D-$3$-branes are wrapped on the $(q_2,p_2)$-cycle the quantisation
condition on the angle $\z$ is
\eqn
\cot{\z}=\frac{q_2}{p_2}\frac{R_4}{R_5}\ .
\label{2.3}
\eeqn
Thus, the moduli $R_4$ and $R_5$ are not independent but obey
the relation
\eqn
\left(\frac{R_4}{R_5}\right)^2=
\frac{p_1}{q_1}\frac{p_2}{q_2}\ .
\label{2.4}
\eeqn

The mass formula for $N$ D-$p$-branes wrapped on a $p$-torus is given by
\cite{Polc1}
\eqn
M_p=N\frac{\left(\o^1R_1\right)...\left(\o^pR_p\right)}
{g\a'^{\frac{p+1}{2}}}\ ,
\label{2.5}
\eeqn
where $g$ is the string coupling constant, $2\pi\alpha'$ is the
inverse string
tension and $R_i$ is the radius of the circle along which the D-branes
are wrapped with winding number $\o^i$. It is straightforward to
generalise this formula for the
case where a given direction of the D-branes wraps around a
$2$-torus. For the configuration described by (\ref{2.1}) we have
\eqna
M_{3_1}=N_1\frac{R_1R_2}{g\a'^2}
\sqrt{\left(q_1R_4\right)^2+\left(p_1R_5\right)^2}\ ,
\nonumber
\\\\
M_{3_2}=N_2\frac{R_1R_3}{g\a'^2}
\sqrt{\left(q_2R_4\right)^2+\left(p_2R_5\right)^2}\ .
\nonumber
\label{2.6}
\eeqna

We proceed by performing a $T$-duality transformation along the
$x^4$-direction on the configuration (\ref{2.1}). The resulting
D-brane system is described by \cite{CostaCvet}
\eqn
\arr{rl}
4_1:&X^1,X^2,X^4,X^5,\\
    &F_{45}^{(1)}=\tan{\z}\ ,\\
4_2:&X^1,X^3,X^4,X^5,\\
    &F_{45}^{(2)}=-\cot{\z}\ .
\earr
\label{2.7}
\eeqn
Consider the first group of D-$4$-branes. They arise by performing a
$T$-duality transformation along the $x^4$-direction on $N_1$
D-$3$-branes that were wrapped on the $x^5$-direction with winding
number $p_1$. Therefore we end up with $N_1$ D-$4$-branes with
winding $p_1$ in the $x^5$-direction. Also, we have seen that the
D-$3$-branes were wrapped on the dualised direction with winding
$-q_1$. Therefore the D-$4$-branes will also carry D-$2$-brane
charge. They carry $-N_1q_1$ units of such charge. In the language of
\cite{Polc1}, the D-$2$-branes dissolve in the D-$4$-branes leaving
flux. This fact is described by the magnetic field condensate
$F^{(1)}$. Similar comments apply to the second group of
D-$4$-branes. The resulting intersecting D-$4$-brane system does not
obey the standard intersecting rules, but supersymmetry of this
configuration is guaranteed by the presence of non-vanishing
worldvolume gauge fields \cite{BalaLeig}.

In the dual theory the quantisation conditions on the rotation angle
$\z$ translate into quantisation conditions on the flux carried by the
worldvolume gauge field strengths $F^{(1)}$ and $F^{(2)}$. In terms of the
dual radius $\left(R_4 \rightarrow \frac{\a'}{R_4}\right)$ and dual string
coupling $\left( g \rightarrow g\frac{\sqrt{\a'}}{R_4}\right) $ these
read as
\eqna
&&F_{45}^{(1)}=\tan{\z}=
\frac{q_1}{p_1}\frac{\a'}{R_4R_5}\ ,
\nonumber
\\\label{2.8}\\
&&F_{45}^{(2)}=-\cot{\z}=
-\frac{q_2}{p_2}\frac{\a'}{R_4R_5}\ .
\nonumber
\eeqna
Similarly, the mass of the D-$4$-branes is now given by
\eqna
M_{4_1}=N_1\frac{R_1R_2}{g\a'^{5/2}}
\sqrt{\left(q_1\a'\right)^2+\left(p_1R_4R_5\right)^2}
=p_1N_1\frac{R_1R_2R_4R_5}{g\a'^{5/2}}
\sqrt{1+\left(F_{45}^{(1)}\right)^2}\ ,
\nonumber
\\\label{2.9}\\
M_{4_2}=N_2\frac{R_1R_3}{g\a'^{5/2}}
\sqrt{\left(q_2\a'\right)^2+\left(p_2R_4R_5\right)^2}
=p_2N_2\frac{R_1R_3R_4R_5}{g\a'^{5/2}}
\sqrt{1+\left(F_{45}^{(2)}\right)^2}\ .
\nonumber
\eeqna
The factor $\sqrt{1+F^2}$ multiplying the usual D-$4$-brane mass
formula is due to the presence of the magnetic field condensate. As
it will be explained in the subsection 3.4, this term also arises in
the vacuum amplitude calculation for a  system of two parallel
D-branes with such fields.

Finally, we remark that all constituent D-branes in the configuration
(\ref{2.7}), including the D-$2$-branes that are dissolved in the
D-$4$-branes, have a common $x^1$-direction. As a consequence, the
D-branes may carry momentum along this direction while leaving some
unbroken supersymmetry. If we introduce $N$ units of (left-moving)
momentum, the compactified system will have an extra mass term given
by
\eqn
P=M_P=\frac{N}{R_1}\ .
\label{2.10}
\eeqn

\subsection{Supergravity solution}

The D-brane system presented above has a corresponding
supergravity solution. This solution may be found by applying
$T$-duality at an angle on the $3\perp 3$ D-brane solution
or by using the $SL(2,\bR)$ symmetry of $N=2$, $D=8$ supergravity
\cite{Izqu..,CostaCvet}. The corresponding type IIA background
fields are
\eqna
ds^{2}&=&\left( H_1H_2\right)^{\frac{1}{2}}\left[ (H_1H_2)^{-1}
\left( -dt^2 +dx_1^2 +(H_3^{\phantom{\tilde{1}}}-1)(dx_1-dt)^2\right)
\right. \nonumber\\\nonumber\\
&&\left. +H_1^{-1}dx_2^2+H_2^{-1}dx_3^2+
\tilde{H}_{12}^{-1}\left(dx_4^2+dx_5^2\right)+ds^2(\bE^4)\right]\
,\nonumber\\\nonumber\\
e^{2(\phi-\phi_{\infty})}&=& 
\tilde{H}_{12}^{-1}\left( H_1H_2\right)^{\frac{1}{2}}\ ,
\nonumber\\\label{2.11}\\
{\cal F}&=&-\sin{\z}\left[dH_1^{-1}\w dt\w dx_1\w dx_2 
+ \star dH_2\w dx_2\right]\nonumber\\\nonumber\\ 
&&-\cos{\z}\left[ \star dH_1\w dx_3 
+dH_2^{-1}\w dt\w dx_1\w dx_3 \right]\ ,\nonumber\\\nonumber\\
{\cal H}&=&-\cos{\z}\sin{\z}\ 
d\left(\frac{H_2-H_1}{\tilde{H}_{12}}\right)\w dx_4\w dx_5\ ,
\nonumber
\eeqna
where
\eqn
H_i=1+\frac{\a_i}{r^2}\ ,\ \ \ \ \ i=1,2,3,
\label{2.12}
\eeqn
are one-centered harmonic function on the $\bE^4$ Euclidean space and
$\star$ is the dual operator with respect to the Euclidean metric on
$\bE^4$. The harmonic function $\tilde{H}_{12}$ is defined by
$\tilde{H}_{12}= \cos^2{\z}\ H_1+\sin^2{\z}\ H_2$. Reducing the solution
(\ref{2.11}) to five dimensions we obtain a black hole with a regular
horizon. The corresponding Einstein frame metric is
\eqn
ds_E^2=-\left(H_1H_2H_3\right)^{-\frac{2}{3}}dt^2
+\left(H_1H_2H_3\right)^{\frac{1}{3}}ds^2(\bE^4)\ .
\label{2.13}
\eeqn
Calculating the horizon area the Bekenstein-Hawking entropy is seen to
be
\eqn
S_{BH}=\frac{A_H}{4G_N^5}=\frac{A_3}{4G_N^5}\sqrt{\a_1\a_2\a_3}\ ,
\label{2.14}
\eeqn
where $G_N^5$ is the five-dimensional Newton's constant and
$A_3=2\pi^2$ is the volume of the unit $3$-sphere.

We want to express the macroscopic entropy formula (\ref{2.14}) in
terms of the quantities describing the D-brane system
(\ref{2.7}). This may be done by matching the mass formulae
(\ref{2.9}) and (\ref{2.10}) to the corresponding string frame
five-dimensional masses \cite{CallMald,Mald}. The result is 
\eqn
\frac{16\pi G_N^5}{A_3}M_i=2\a_i\ .
\label{2.15}
\eeqn
The five- and ten-dimensional Newton's constants are related by
\eqn
G_N^{10}=(2\pi)^5R_1...R_5G_N^5=8\pi^6g^2\alpha'^4\ .
\eeqn 
After some straightforward algebra the previous relations will give
the following entropy formula
\eqn
S_{BH}=2\pi\sqrt{NN_1N_2n_L}\ ,
\label{2.16}
\eeqn
where
\eqn
n_L=p_2q_1+p_1q_2\ ,
\label{2.17}
\eeqn
is, as we shall see, the Landau degeneracy for open strings with ends
on different branes. In the T-dual picture described by (\ref{2.1})
the same factor arises in the entropy formula. This factor is
interpreted as the D-3-branes' number of intersecting points on the
$x^4x^5$ 2-torus \cite{Bala..}. The string theory entropy counting is
then clear because on the intersecting points strings with both ends
on different D-3-branes become massless.

The D-$4$-brane charges $Q_i=N_ip_i$ may be expressed as an
integral over spacial infinity of the $4$-form field strength. The
properly normalised expression for these charges is
\eqn
Q_i=\frac{1}{2A_3(2\pi)g\a'^{3/2}}\int_{\Sigma_i}{\cal F}
=\left\{
\arr{l}
\frac{R_3}{g\a'^{3/2}}\cos{\z}\ \a_1\ ,\ \ \ \ i=1,\\\\
\frac{R_2}{g\a'^{3/2}}\sin{\z}\ \a_2\ ,\ \ \ \ i=2,
\earr
\right.
\label{2.18}
\eeqn
where $\Sigma_i=S^3_{\infty}\times S^1_i$ with $S^3_{\infty}$ the
asymptotic $3$-sphere and $S^1_i$ the compactified dimension
transverse to the $i$-th $4$-branes. We remark that it is also possible
to write the D-$2$-brane charges carried by the D-$4$-branes as an
integral of the dual of the $4$-form field strength. The `charge'
associated with the (left-moving) momentum carried by the D-$4$-branes
may be calculated by using equations (\ref{2.10}) and (\ref{2.15})
\eqn
N=\frac{R_1^2R_2R_3R_4R_5}{g^2\a'^4}\ \a_3\ .
\label{2.19}
\eeqn

\subsection{Non-extreme solution}

The supergravity solution (\ref{2.11}) can be easily generalised to the
corresponding non-extreme solution \cite{Duff..,CvetTsey,Costa}. Start
with the solution (\ref{2.11})
without the plane wave and perform the following redefinitions
\eqn
\arr{rl}
dt^2\rightarrow & f\ dt^2\ ,\\
dr^2\rightarrow & f^{-1}dr^2\ ,\\
Q_i\rightarrow & Q_i\sqrt{(\mu +\a_i)/\a_i}\ ,\ \ \ \ \ i=1,2,
\earr
\label{2.20}
\eeqn
where $f=1-\frac{\mu}{r^2}$ and $Q_i$ is either the electric or
magnetic charge associated with the $i$-th $4$-branes. For later
convenience we will express our solution in terms of the parameters
$r_0$ and $\b_i$ defined by
\eqn
\arr{l}
\mu=r_0^2\ ,\\
\a_i=r_0^2\sinh^2{\b_i}\ ,\ \ \ \ \ i=1,2.
\earr
\label{2.21}
\eeqn
In the limit $r_0 \rightarrow 0$ and $\b_i \rightarrow \infty$,
holding $\alpha_i$ fixed, we obtain the extreme solution. The charge
associated with the momentum along the $x^1$-direction may be obtained
by performing the boost transformation
\eqn
\arr{rl}
t\rightarrow &\cosh{\b_3}\ t-\sinh{\b_3}\ x^1\ ,\\
x^1\rightarrow &-\sinh{\b_3}\ t+\cosh{\b_3}\ x^1\ .
\earr
\label{2.22}
\eeqn
The resulting non-extreme solution is described by
\eqna
ds^{2}&=&\left( H_1H_2\right)^{\frac{1}{2}}\left[ (H_1H_2)^{-1}
\left( -dt^2 +dx_1^2 +\left(\frac{r_0}{r}\right)^2
\left(\cosh{\b_3}\ dt-\sinh{\b_3}\ dx_1\right)^2\right)\right. 
\nonumber\\\nonumber \\
&&\left. +H_1^{-1}dx_2^2+H_2^{-1}dx_3^2+
\tilde{H}_{12}^{-1}\left(dx_4^2+dx_5^2\right)
+f^{-1}_{\phantom{\frac{1}{2}}}dr^2+r^2d\Omega_3^2\right]\ ,
\nonumber\\\label{2.23}\\
{\cal F}&=&-\sin{\z}\left[\coth{\b_1}\ dH_1^{-1}\w dt\w dx_1\w dx_2 
+\coth{\b_2}\ \star dH_2\w dx_2\right]
\nonumber\\\nonumber\\ 
&&-\cos{\z}\left[ \coth{\b_1}\ \star dH_1\w dx_3 
+\coth{\b_2}\ dH_2^{-1}\w dt\w dx_1\w dx_3 \right]\ ,
\nonumber
\eeqna
and the dilaton field and $3$-form field strength are still given as in
(\ref{2.11}). The charges (\ref{2.18}) and (\ref{2.19}) previously
defined are now given by
\eqna
Q_1&=&\frac{R_3}{2g\a'^{3/2}}\cos{\z}\ r_0^2\ \sinh{2\b_1}\ ,
\nonumber\\\nonumber\\
Q_2&=&\frac{R_2}{2g\a'^{3/2}}\sin{\z}\ r_0^2\ \sinh{2\b_2}\ ,
\label{2.24}
\\\nonumber\\
N&=&\frac{R_1^2R_2R_3R_4R_5}{2g^2\a'^4}\ r_0^2\ \sinh{2\b_3}\ .
\nonumber
\eeqna

The ADM mass, entropy and Hawking temperature may be calculated by
reducing our solution to five dimensions. The result is
\eqna
M&=&\frac{R_1R_2R_3R_4R_5}{2g^2\a'^4}r_0^2\ 
\sum_{i=1}^{3}\cosh{2\b_i}\ ,
\nonumber\\\nonumber\\
S_{BH}&=&2\pi\frac{R_1R_2R_3R_4R_5}{g^2\a'^4}r_0^3\ 
\prod_{i=1}^3\cosh{\b_i}\ ,
\label{2.25}
\\\nonumber\\
T_H&=&\left( 2\pi r_0\prod_{i=1}^3\cosh{\b_i}\right)^{-1}
=S_{BH}^{-1}\frac{r_0^2R_1R_2R_3R_4R_5}{g^2\a'^4}\ .
\nonumber
\eeqna
As in the extreme case, we are interested in expressing these
quantities in terms of an associated  D-brane system. In other words,
we want to introduce right-moving momentum and anti-D-4-branes. In the
near-extremal limit the resulting system will describe the
near-extremal black-hole \cite{Horo..}.
With that in mind, we define the 
quantities ($N_1$, $N_{\bar{1}}$, $N_2$, $N_{\bar{2}}$, $N_R$, $N_L$) by
\eqn
\arr{ll}
Q_{4_1}=
N_1p_1=\frac{R_3r_0^2}{4g\a'^{3/2}}\cos{\z}\ e^{2\b_1}\ ,\ \ \ \ &
Q_{\bar{4}_1}=
N_{\bar{1}}p_1=\frac{R_3r_0^2}{4g\a'^{3/2}}\cos{\z}\ e^{-2\b_1}\ ,
\\\\
Q_{4_2}=
N_2p_2=\frac{R_2r_0^2}{4g\a'^{3/2}}\sin{\z}\ e^{2\b_2}\ ,&
Q_{\bar{4}_2}=
N_{\bar{2}}p_2=\frac{R_2r_0^2}{4g\a'^{3/2}}\sin{\z}\ e^{-2\b_2}\ ,
\\\\
N_L=
\frac{R_1^2R_2R_3R_4R_5r_0^2}{4g^2\a'^4}\ e^{2\b_3}\ ,&
N_R=
\frac{R_1^2R_2R_3R_4R_5r_0^2}{4g^2\a'^4}\ e^{-2\b_3}\ .
\earr
\label{2.26}
\eeqn
We shall see in section 4  that in the near-extremal limit, these
quantities are interpreted as the units
of left ($N_L$) and right ($N_R$) moving momentum along the
$x^1$-direction, and as the winding numbers along the $x^1$-direction of the
D-$4$-branes ($N_1$, $N_2$) and anti-D-$4$-branes ($N_{\bar{1}}$,
$N_{\bar{2}}$) both with the
worldvolume gauge field turned on as in (\ref{2.7}). In
terms of these quantities the ADM mass, entropy and Hawking
temperature in (\ref{2.25}) are given by
\eqna
M&=&
\frac{R_1R_2R_4R_5}{g\a'^{5/2}}p_1\left( N_1+N_{\bar{1}}\right)
\sqrt{1+\left( F_{45}^{(1)}\right)^2} \nonumber\\\nonumber\\ &&
+\frac{R_1R_3R_4R_5}{g\a'^{5/2}}p_2\left( N_2+N_{\bar{2}}\right)
\sqrt{1+\left( F_{45}^{(2)}\right)^2}+
\frac{1}{R_1}\left( N_R+N_L\right)\ ,
\nonumber\\\label{2.28}\\
S_{BH}&=&
2\pi\left(\sqrt{N_1}+\sqrt{N_{\bar{1}}}\right)
\left(\sqrt{N_2}+\sqrt{N_{\bar{2}}}\right)
\left(\sqrt{N_R}+\sqrt{N_L}\right) \sqrt{n_L}\ ,\nonumber\\\nonumber\\
T_H&=&
S_{BH}^{-1}\left( \frac{4}{R_1}\right) \sqrt{N_RN_L}\ ,\nonumber
\eeqna
where $n_L$ is defined in (\ref{2.17}).

\section{D-branes with worldvolume magnetic fields}
\news

In this section we will study open superstrings ending on
D-branes which have a worldvolume gauge field strength of the form
\eqn
F_{\ha\hb}=\left(
\arr{cc}
0&f\\
-f&0
\earr
\right)
\label{3.1}
\eeqn
where $f$ is a constant and the indices $\ha$ and $\hb$ label two spatial
worldvolume coordinates.\footnote{D-branes with worldvolume field
  condensates have been earlier considered in
  \cite{Li,CallKleb,GreenGutp} using the boundary state formalism.} We
will just consider the
superstring mode expansion for these two coordinates. This, together
with well known facts, will be sufficient to explain the three open
string sectors of the D-brane system (\ref{2.7}), i.e. the two sectors
corresponding to open strings with both ends on the same D-branes and
the sector corresponding to open strings with each end on different 
D-branes. In the first cases, the field strength (\ref{3.1}) is equal
on both ends of the string while in the second case it is not. Of
course there are other possibilities, for example one end of the
string may be let to move freely \cite{Lifs}. An
important remark is that we are just
considering the $U(1)$ center of the D-branes' worldvolume
fields. In our D-brane system we have $N_1$ and $N_2$ D-4-branes on
top of each other wrapping on the $x^5$-direction with winding numbers
$p_1$ and $p_2$, respectively. This means that the corresponding
worldvolume fields are in the adjoint representation of
$U(N_ip_i)$. As explained in the introduction, the $U(1)$ flux induces
a twist on the fields. The gauge potential as well as the scalar
fields will then obey twisted boundary conditions
\cite{tHoo,tHoo1,GuraRamg,HashTayl}. We can choose a set of
multiple transition functions on this $U(N_ip_i)$ bundle such that the
only non-vanishing D-branes' worldvolume fields are in the $U(1)$
center of the group.
This means that we can perform the calculations as if our open
strings have each of its ends on a singly wrapped D-brane.

Let us start by writing the boundary term of the open superstring
action in the presence of a $U(1)$ background gauge field. Using the
orthonormal gauge coordinates $\t$ $(-\infty<\t<\infty)$ and
$\s$ $(0<\s<\pi)$ we have (taking $\a'=1/2$)
\eqn
S_b=\frac{1}{\pi}\left.\int_{-\infty}^{+\infty} d\t\left[
A_a\p_\t X^a
-\frac{i}{4}F_{ab}\left(\Psi_-^a\Psi_-^b+\Psi_+^a\Psi_+^b\right)
\right]\right|_{\s =0}^{\s =\pi}\ ,
\label{3.2}
\eeqn
where $a,b=0,...,9,$ and $\Psi_-^a$ and $\Psi_+^a$ are
respectively the positive and negative chirality spinor components of
the worldsheet Majorana spinor $\Psi^a$ \cite{Green..}. Performing $T$-duality
transformations along the directions $m=p+1,...,9$ we find the
boundary term for the action describing open superstrings with the
ends on a D-$p$-brane with worldvolume fields $A_{\a}(X^{\b})$
($\a,\b=0,...,p$) and $\phi^m(X^{\a})$
\eqna
S_b&=&\frac{1}{\pi}\int_{-\infty}^{+\infty} d\t
\left[ \left( A_{\a}\p_\t X^{\a} +\phi_m\p_\s X^m\right)
\phantom{\frac{i}{4}}\right.\nonumber\\\label{3.3}\\
&& \left.\left.-\frac{i}{4} F_{\a\b}\left(\Psi_-^{\a}\Psi_-^{\b}+
\Psi_+^{\a}\Psi_+^{\b}\right)
+\frac{i}{2}\p_{\a}\phi_m\left(\Psi_-^{\a}\Psi_-^m-
\Psi_+^{\a}\Psi_+^m\right)
\right]\right|_{\s =0}^{\s =\pi}\ ,\nonumber
\eeqna
where we used the $T$-duality transformation rules $\p_{\t}
\rightarrow \p_{\s}$, $A_m \rightarrow \phi_m$,
$\Psi_{\mp}^m \rightarrow \mp\Psi_{\mp}^m$. If the open string
ends on different D-branes the background fields will differ for
$\s=\pi$ and $\s=0$.

As explained above, we will find the superstring
mode expansion for the case where the only non-vanishing worldvolume
fields are
\eqn
\arr{l}
F_{\ha\hb}^{(1)}\equiv F_{\ha\hb}(\s =0)=\tan{\g}\left(
\arr{cc}
0&1\\
-1&0
\earr
\right)\ ,
\\\\
F_{\ha\hb}^{(2)}\equiv F_{\ha\hb}(\s =\pi)=\tan{\g'}\left(
\arr{cc}
0&1\\
-1&0
\earr
\right)\ ,
\earr
\label{3.4}
\eeqn
where $\g$ and $\g'$ are arbitrary constants and the indices
$\ha$ and $\hb$ label only two spatial worldvolume coordinates. For
the D-brane system considered in this paper, and for the open string
sector corresponding to open strings ending on different branes, the
indices $\ha$ and $\hb$ label the $x^4$- and $x^5$-directions and
\eqn
\tan{\g}=\frac{q_1}{p_1}\frac{\a'}{R_4R_5}\ ,\ \ \ \ \ 
\tan{\g'}=-\frac{q_2}{p_2}\frac{\a'}{R_4R_5}\ .
\label{3.4a}
\eeqn

The variation of the $X^{\ha}$ bosonic fields in the superstring
action with a
boundary term like in (\ref{3.3}) yields the usual wave equation
together with the boundary conditions
\eqn
\p_{\s}X^{\ha}=
F^{\ha}_{\ \hb}\p_{\t}X^{\hb}\ ,\ \ \ \ \ 
\s=0,\pi.
\label{3.5}
\eeqn
Similarly, to cancel the boundary term that arises on 
varying the fermionic fields in the action (\ref{3.3}), the right
($\Psi_-^{\ha}$) and left ($\Psi_+^{\ha}$) moving world-sheet
fermions should obey the boundary conditions \cite{Call..,BachPorr}
\eqn
\arr{c}
\Psi_+^{\ha}-\Psi_-^{\ha}=F^{\ha}_{\ \hb}
\left( \Psi_+^{\hb}+\Psi_-^{\hb}\right) \ ,\ \ \ \ \ \s =0,
\\\\
\Psi_+^{\ha}\mp\Psi_-^{\ha}=F^{\ha}_{\ \hb}
\left( \Psi_+^{\hb}\pm\Psi_-^{\hb}\right) \ ,\ \ \ \ \ \s =\pi.
\earr
\label{3.6}
\eeqn
The upper and lower sign choices correspond to the
Ramond  and Neveu-Schwarz sectors of the theory, respectively. In
fact, the boundary conditions (\ref{3.6}) are found by requiring the
fermionic fields variations to obey 
$\d \Psi^{\ha}_+=\d \Psi^{\ha}_-$
at $\s=0$, and $\d\Psi_+^{\ha}=\pm \d \Psi^{\ha}_-$
at $\s=\pi$.

\subsection{Bosonic mode expansion}

In this subsection we will study the mode expansion for the worldsheet
bosonic fields satisfying the boundary conditions (\ref{3.5}). All the
results presented here have been obtained in \cite{Abou..} in the
context of open strings in a constant magnetic field, and are included
for the sake of clarity. Has explained
above we will just consider the excitations along the
$x^{\ha}$-directions. Relabelling these worldsheet fields by $X$
and $Y$ we define the complex field
\eqn
Z\equiv\frac{1}{\sqrt{2}}\left( X+iY\right)\ .
\label{3.7}
\eeqn
The boundary conditions (\ref{3.5}) now read
\eqn
\arr{ll}
\p_{\s}Z=-i\tan{\g}\ \p_{\t}Z\ ,\ \ \ &\s =0,\\
\p_{\s}Z=-i\tan{\g'}\ \p_{\t}Z\ ,&\s =\pi.
\earr
\label{3.8}
\eeqn
The mode expansion is then seen to be
\eqn
Z=z+i\left[ \sum_{n=1}^{\infty}a_{n-\e}\phi_{n-\e}(\t ,\s )
-\sum_{n=0}^{\infty}a_{n+\e}^{\dagger}\phi_{-n-\e}(\t ,\s )\right]\ ,
\label{3.9}
\eeqn
where
\eqn
\phi_{n-\e} =\frac{1}{\sqrt{|n-\e|}}\cos{[(n-\e)\s+\g]}e^{-i(n-\e)\t}\ ,
\label{3.10}
\eeqn
with $\e=(\g-\g')/\pi$, $n$ an integer and
$a_{n-\e},a_{n+\e}^{\dagger}$ real. Note that by an
appropriate redefinition of the modes in (\ref{3.9}) the parameter
$\e$ may be taken to lie between $0$ and $1$. In fact, by
interchanging the $a_{n-\e}$ and $a_{n+\e}$ mode operators it may be
taken to lie between 0 and 1/2. Since the
integrals of the mode functions $\phi_{n-\e}$ are non-zero, the constant
$z$ can not be interpreted as the string's center-of-mass. The functions
$\phi_{n-\e}$, together with a constant function, form a complete basis of
functions on the interval $[0,\pi ]$ (the details may be found in
\cite{Abou..}).

The canonical momentum is as usual given by
\eqn
P_Z=\frac{\p {\cal L}}{\p \left(\p_{\t}Z\right) }=\frac{1}{\pi}
\left[ \p_{\t}Z^{\dagger} + 
A^{(2)\dagger}\d (\pi -\s ) - A^{(1)\dagger}\d (\s ) \right]\ ,
\label{3.11}
\eeqn
where the complex gauge fields $A^{(i)}$ are defined by
\eqn
A^{(i)}=\frac{1}{\sqrt{2}}
\left( A_x^{(i)}+iA_y^{(i)}\right)\ ,\ \ \ i=1,2.
\label{3.12}
\eeqn
At this point we have to choose a gauge for the D-branes'
worldvolume gauge fields. A convenient choice is
\eqna
A^{(1)}_{\ha}dx^{\ha}=
\frac{\tan{\g}}{2}\left( Xdy-Ydx\right)\ ,\nonumber\\\label{3.13}\\
A^{(2)}_{\ha}dx^{\ha}=
\frac{\tan{\g'}}{2}\left( Xdy-Ydx\right)\ .\nonumber
\eeqna
The canonical momentum is then seen to be
\eqn
P_Z=\frac{1}{\pi}\left( \p_{\t}Z^{\dagger}+
\frac{i}{2}Z^{\dagger}\left[\tan{\g}\ \d (\s )
-\tan{\g'}\ \d (\pi-\s )\right]\right)\ .
\label{3.14}
\eeqn
The first quantised string is obtained by introducing the equal time
canonical commutation relations. This gives the following
non-vanishing commutators for the mode operators in the expansion
(\ref{3.9})
\eqn
\left[ a_{n-\e},a_{m-\e}^{\dagger}\right] =\d _{nm}\ ,\ \ \ 
\left[ a_{n+\e},a_{m+\e}^{\dagger}\right] =\d _{nm}\ ,\ \ \ 
\left[ z,z^{\dagger}\right] =\frac{\pi}{\tan{\g}-\tan{\g'}}\ .
\label{3.15}
\eeqn
Note that these commutation relations do not depend on the gauge choice
(\ref{3.13}). Thus, the spectrum of string states is gauge
invariant. 

Introducing the light cone coordinates $\s^{\pm}=\t \pm \s$,
the Virasoro operators are defined to be
\eqn
L_n=\frac{1}{\pi}\int_0^{\pi}d\s\left( e^{in\s}T_{++}+
e^{-in\s}T_{--}\right)\ ,
\label{3.16}
\eeqn
where $T_{\pm\pm}$ are the $\pm\pm$ components of the energy-momentum
tensor. The corresponding contributions from the $X$ and $Y$
worldsheet fields are
\eqn
T_{\pm\pm}=\p_{\pm}X^{\ha}\p_{\pm}X_{\ha}=
\frac{1}{2}\left( \p_{\t}Z\pm\p_{\s}Z\right)
\left( \p_{\t}Z\pm\p_{\s}Z\right)^{\dagger}\ .
\label{3.17}
\eeqn
After some straightforward algebra we obtain for $n\ge 0$
\eqna
L_n^{(B)}&=&\sum_{p=1}^{\infty}\sqrt{(p-\e)(n+p-\e)}\
a_{p-\e}^{\dagger}a_{n+p-\e}\nonumber\\\nonumber\\ &&
+\sum_{p=0}^{\infty}\sqrt{(p+\e)(n+p+\e)}\ a_{p+\e}^{\dagger}a_{n+p+\e}
\label{3.18}
\\\nonumber\\ &&
+\sum_{p=0}^{n-1}\sqrt{(p+\e)(n-p-\e)}\ a_{p+\e}a_{n-p-\e}\ .
\nonumber
\eeqna
For $n<0$, $L_n$ is obtained by substituting $n$ by $-n$ and by taking
the hermitian conjugate of (\ref{3.18}). The Virasoro operators
satisfy the algebra
\eqn
\left[ L_n^{(B)}, L_m^{(B)} \right] =(n-m)
L_{n+m}^{(B)}+A(n)\d_{n+m,0}\ ,
\label{3.19}
\eeqn
where the $c$-number $A(n)$ is due to the usual normal ordering
ambiguity in the definition of $L_0$. The value of $A(n)$ may be
calculated for the bosonic string as it was done in \cite{Abou..}. We
will calculate this anomalous term in the full superstring theory.

As a final remark, we note that the frequencies of the $a_{n-\e}$ and
$a_{n+\e}$ ($n>0$) oscillators will be shifted with respect to their
usual values by $-\e$ and $+\e$, respectively. Also, the operators
$a_{\e}$ and $a_{\e}^{\dagger}$ annihilate and create quanta of frequency
$\e$. If $\e=0$ the zero mode operators disappear (we will
comment on this case in subsection 3.4). If $\e=\frac{1}{2}$ the
frequencies of the mode operators 
$a_{n-\e}$ and $a_{n+\e}$, and therefore the spectrum,  become the
same as those for open
strings with ND boundary conditions. The only
difference is that the zero mode $z$ is not absent.

\subsection{Fermionic mode expansion}

At the beginning of this section we wrote the boundary conditions
satisfied by the worldsheet fermionic fields for the case when the
D-brane worldvolume gauge field does not vanish. As in the bosonic case we
will just consider the two relevant directions corresponding to the
ansatz (\ref{3.4}). Start by defining the complex spinor field
\eqn
\Psi \equiv\frac{1}{\sqrt{2}}\left(\Psi^x+i\Psi^y\right) =
\frac{1}{\sqrt{2}} \left(
\arr{l}
\Psi_-^x+i\Psi_-^y\\
\Psi_+^x+i\Psi_+^y
\earr
\right) \equiv \left(
\arr{l}
\Psi_-\\
\Psi_+
\earr
\right)\ .
\label{3.25}
\eeqn
In terms of this field the boundary conditions (\ref{3.6}) read
\eqn
\arr{ll}
\Psi_+ -\Psi_-=-i\tan{\g}\ \left(\Psi_+ +\Psi_-\right)\ ,\ \ \ &\s=0,\\
\Psi_+\mp\Psi_-=-i\tan{\g'}\ \left(\Psi_+\pm\Psi_-\right)\ ,&\s=\pi,
\earr
\label{3.26}
\eeqn
where the upper and lower sign choices correspond to the Ramond and
Neveu-Schwarz sectors, respectively.

Let us consider first the Ramond sector. The mode expansion for the
field $\Psi$ may be written as
\eqn
\Psi_{\mp}=\frac{1}{\sqrt{2}}\left[
\sum_{n=1}^{\infty}\Psi_{n-\e}\phi_{n-\e}^{(\mp)}(\t ,\s )-
\sum_{n=0}^{\infty}\Psi_{n+\e}^{\dagger}\phi_{-n-\e}^{(\mp)}(\t ,\s )\right]\ ,
\label{3.27}
\eeqn
where
\eqn
\phi_{n-\e}^{(\mp)}=e^{(n-\e )(\t\mp\s )\mp\g }\ ,
\label{3.28}
\eeqn
with $\e$ defined as in (\ref{3.10}) and $n$ an
integer. The canonical momentum of  the complex spinor field $\Psi$ is
\eqn
P_{\Psi}=\frac{\p {\cal L}}{\p \left(\p_{\t}\Psi\right) }
=\frac{i}{2\pi}\Psi^{\dagger}\ .
\label{3.29}
\eeqn
Introducing the equal time anti-commutation relations we find the
following non-vanishing anti-commutators for the mode operators in the
expansion (\ref{3.27})
\eqn
\{\Psi_{n-\e},\Psi_{m-\e}^{\dagger}\}=\d_{nm}\ ,\ \ \ 
\{\Psi_{n+\e},\Psi_{m+\e}^{\dagger}\}=\d_{nm}\ .
\label{3.30}
\eeqn
The contribution of the worldsheet fermionic fields $\Psi^x$ and
$\Psi^y$ to the $\pm\pm$ components of the energy-momentum
tensor is
\eqn
T_{\pm\pm}=\frac{i}{2}\Psi_{\pm}^{\ha}\p_{\pm}\Psi_{\pm\ha}=
\frac{i}{2}\left(\Psi_{\pm}\p_{\pm}\Psi_{\pm}^{\dagger}+
\Psi_{\pm}^{\dagger}\p_{\pm}\Psi_{\pm}\right)\ .
\label{3.31}
\eeqn
Thus, using the definition (\ref{3.16}) and after some algebra we find
that for $n \ge 0$
\eqna
L_n^{(R)}&=&
L_n^{(B)}+\sum_{p=1}^{\infty}\left(\frac{n}{2}+p-\e\right)
\ \Psi_{p-\e}^{\dagger}\Psi_{n+p-\e}\nonumber\\\nonumber\\ &&
+\sum_{p=0}^{\infty}\left(\frac{n}{2}+p+\e\right)
\ \Psi_{p+\e}^{\dagger}\Psi_{n+p+\e}
\label{3.32}
\\\nonumber\\ &&
+\sum_{p=0}^{n-1}\left(\frac{n}{2}-p-\e\right)\ \Psi_{n-p-\e}\Psi_{p+\e}\ ,
\nonumber
\eeqna
where $L_n^{(B)}$ is the bosonic contribution to the Virasoro operators
given in (\ref{3.18}). For $n<0$ we have to replace $n$ by $-n$ and
take the hermitian conjugate. The Virasoro algebra may be seen to be
\eqn
\left[ L_n^{(R)}, L_m^{(R)} \right] =(n-m)
L_{n+m}^{(R)}+\frac{2}{8}n^3\d_{n+m,0}\ .
\label{3.33}
\eeqn
This is exactly the standard result expected for the Ramond
sector. Thus the zero energy is not changed in the presence of the
magnetic field condensate (\ref{3.4}). This is just the consequence
of the fact that both the worldsheet bosons and fermions have the same
moding. As in the bosonic case the frequencies of the $\Psi_{n-\e}$
and $\Psi_{n+\e}$
($n>0$) oscillators are shifted with respect to their usual values by
$-\e$ and $\e$, respectively. Also, the operators $\Psi_{\e}$ and
$\Psi_{\e}^{\dagger}$ annihilate and create quanta of frequency $\e$.
If $\e=0$ these operators will take ground states into ground states
forming the usual Ramond vacua. If $\e=1/2$ the mode operators,
and therefore the spectrum, become similar to the case of open
superstrings with ND boundary conditions, i.e. the expansion becomes
of Neveu-Schwarz type.

We now turn to the Neveu-Schwarz sector of the theory. In this case
the mode expansion takes the form
\eqn
\Psi_{\mp}=\frac{1}{\sqrt{2}}\left[
\sum_{r=\frac{1}{2}}^{\infty}\Psi_{r-\e}
\phi_{r-\e}^{(\mp)}(\t ,\s )-
\sum_{r=\frac{1}{2}}^{\infty}\Psi_{r+\e}^{\dagger}
\phi_{-r-\e}^{(\mp)}(\t ,\s )\right]\ ,
\label{3.34}
\eeqn
where
\eqn
\phi_{r-\e}^{(\mp)}=e^{(r-\e )(\t\mp\s )\mp\g }\ ,
\label{3.35}
\eeqn
with $\e$ defined as before and $r$ a
half-integer. After quantisation the non-vanishing anti-commutation
relations for the mode operators in the expansion (\ref{3.34}) are
\eqn
\{\Psi_{r-\e},\Psi_{s-\e}^{\dagger}\}=\d_{rs}\ ,\ \ \ 
\{\Psi_{r+\e},\Psi_{s+\e}^{\dagger}\}=\d_{rs}\ .
\label{3.36}
\eeqn
A similar calculation to the one in the previous case gives the
Virasoro operators ($n \ge 0$)
\eqna
L_n^{(NS)}&=&
L_n^{(B)}+\sum_{r=\frac{1}{2}}^{\infty}\left(\frac{n}{2}+r-\e\right)
\ \Psi_{r-\e}^{\dagger}\Psi_{n+r-\e}\nonumber\\\nonumber\\ &&
+\sum_{r=\frac{1}{2}}^{\infty}\left(\frac{n}{2}+r+\e\right)
\ \Psi_{r+\e}^{\dagger}\Psi_{n+r+\e}
\label{3.37}\\\nonumber\\ &&
+\sum_{r=\frac{1}{2}}^{n-\frac{1}{2}}\left(\frac{n}{2}-r-\e\right)
\ \Psi_{n-r-\e}\Psi_{r+\e}\ .
\nonumber
\eeqna
For $n<0$ we have to replace $n$ by $-n$ and take the hermitian
conjugate. The Virasoro algebra is now given by
\eqn
\left[ L_n^{(NS)}, L_m^{(NS)} \right] =(n-m)
L_{n+m}^{(NS)}+\d_{n+m,0}\left[\frac{2}{8}(n^3-n)+\e n\right]\ .
\label{3.38}
\eeqn
The first term inside the square brackets is the usual term that would
be obtained if the magnetic field was absent. The other term is
proportional to $n$ and it may be absorbed by a redefinition $L_0
\rightarrow L_0+\e/2$, which shifts the zero energy by a term
$\e/2$ (in units of $(\a')^{-1}$).

As in the previous cases the frequencies of the $\Psi_{r-\e}$ and
$\Psi_{r+\e}$ oscillators are shifted with respect to their usual
values by $-\e$ and $\e$, respectively. If $\e =0$ we just obtain the
standard spectrum for the worldsheet fermions in the NS sector. If
$\e =\frac{1}{2}$ the frequency of the mode operators $\Psi_{1/2-\e}$ and
$\Psi_{1/2-\e}^{\dagger}$ vanishes. In this case the spectrum is
similar to open superstrings with ND boundary conditions, i.e. it
becomes of Ramond type.

\subsection{The Landau degeneracy}

In the subsection 3.1 we have seen that the zero modes $z$ and
$z^{\dagger}$ are
non-commuting variables. In terms of the $x$ and $y$ coordinates the
corresponding commutation relation reads
\eqn
\left[ x,y \right] =i\frac{\pi}{\tan{\g} -\tan{\g'}}\ .
\label{3.20}
\eeqn
Thus, $k=y(\tan{\g}-\tan{\g'})/\pi$ may be seen as the conjugate
momentum of $x$. Since both zero modes $x$ and $y$ commute with $L_0$
we can take
any string state to be an eigenstate of either $x$ or $y$. In the
D-brane system we are interested on the $x$- and $y$-directions are
compactified on a torus. As a result, the eigenstates of $x$ or $y$
will have a degeneracy, the Landau degeneracy \cite{Abou..}.

Assume that the D-branes where the open strings end are wrapped around
the $y$-direction with winding numbers $p_1$ and $p_2$
for each D-brane. We consider first the case where $p_1$ and $p_2$ are
co-prime. The open strings ending on both branes carry Chan-Paton
factors in the fundamental representation of $U(p_1)\times
U(\bar{p}_2)$. The worldsheet fields are identified according to 
\eqn
\left( Y(\t,\s)\right)_{a\ov{b}}
+2\pi R_y\equiv 
\left( Y(\t,\s)\right)_{a+1,\ov{b+1}}\ ,
\label{3.21}
\eeqn
where $a=1,...,p_1$ and $\ov{b}=1,...,p_2$. Thus, going around
$p_1p_2$ cycles in the $y$-direction we have the identification
\eqn
\left( Y(\t,\s)\right)_{a\ov{b}} +2\pi p_1p_2R_y\equiv 
\left( Y(\t,\s)\right)_{a\ov{b}}\ .
\label{3.21a}
\eeqn

To deduce the existence of the Landau levels we essentially follow the
same steps as in ref. \cite{Abou..}. Since the D-branes are singly
wrapped in the $x$-direction the zero mode wave function 
$\langle x_{a\ov{b}}|k_{a\ov{b}}\rangle =\exp{(ikx)}$ must be
single-valued. This implies that the $k$ eigenvalues are quantised as 
$(k_m)_{a\ov{b}}=m/R_x$. In terms of the zero mode $y_{a\ov{b}}$
this condition reads
\eqn
(y_m)_{a\ov{b}}=\frac{\pi}{\tan{\g}-\tan{\g'}}\frac{m}{R_x}=
\frac{m}{n_L}L\ ,
\label{3.22}
\eeqn
where $n_L=p_2q_1+p_1q_2$ and $L=2\pi p_1p_2R_y$. In the last equality
we have used equation (\ref{3.4a}) which applies to the D-brane system
considered in this paper. More generally, the fact that $n_L$ has to be an
integer follows from the flux quantisation condition on the D-branes'
gauge fields. This may also be derived has a Dirac quantisation
condition since a constant magnetic field on a torus is a monopole
field \cite{Abou..}.

In order to interpret equation (\ref{3.22}) we note that $p_1$ and
$p_2$ were taken to be co-prime. This means that for a given
$(y_m)_{a\ov{b}}$ all the other zero modes $(y_m)_{a'\ov{b'}}$ are
determined by (\ref{3.21}). Since the system has periodicity $L$ it
follows from (\ref{3.22}) that there are $n_L$ independent
$|y_m\rangle$ states. Thus, the degeneracy of any string state is
$n_L$, i.e. any string state will be found by acting with creation
operators on the degenerated ground state
\eqn
|y_m\rangle\ ,\ \ \ m=1,...,n_L\ .
\label{3.23}
\eeqn

To relax the assumption that $p_1$ and $p_2$ are co-prime, we just
have to realise that in the general case the system has periodicity
$L'=2\pi lp'_1p'_2R_y$, where $p_1=lp'_1$ and $p_2=lp'_2$ with $p'_1$ and
$p'_2$ co-prime and $l$ an integer. Equation (\ref{3.22}) may
be written as
\eqn
(y_m)_{a\ov{b}}=\frac{m}{n'_L}L'\ ,
\label{3.22a}
\eeqn
where $n'_L=p'_2q_1+p'_1q_2$. The point now is that for a given
$(y_m)_{a\ov{b}}$ not all the other zero modes $(y_m)_{a'\ov{b'}}$
are determined by (\ref{3.21}). A minimal set of zero modes that
determines all the other zero modes is 
\eqn
(y_m)_{a\ov{b}}\ ,\ \ (y_m)_{a+1,\ov{b}}\ ,...,\ \ 
(y_m)_{a+l-1,\ov{b}}\ .
\label{3.24}
\eeqn
Thus, there are $n'_L$ Landau levels but each level is itself $l$
times degenerated. This means that again any string state will be
found by acting with the creation operators on the $n_L$ times
degenerated ground state
\eqn
|y_{m,r}\rangle\ ,\ \ \ m=1,...,n'_L,\ \ \ r=1,...,l\ .
\label{3.23a}
\eeqn

\subsection{The $(p|p-2)$ bound state}

For completeness we will comment on the $(p|p-2)$ D-brane bound
state \cite{Doug}. This configuration is interpreted as a D-$p$-brane
carrying also
the charge of a D-$(p-2)$-brane. As it is well known the excitations of
this D-brane bound state are described by open superstrings with the
following boundary condition for the worldsheet bosons (at $\sigma=0$
and $\sigma =\pi$)
\eqn
\arr{ll}
\p_{\s}X^{\a}=0\ , & \a=0,...,p-2,\\
\p_{\s}X^{\ha}=F^{\ha}_{\ \hb}\p_{\t}X^{\hb},\ \ \ &
\ha,\hb =p-1,p,\\
\p_{\t}X^m=0\ , & m=p+1,...,9,
\earr
\label{3.39}
\eeqn
where $F^{\ha}_{\ \hb}$ is given by (\ref{3.1}). The boundary
conditions for the fermionic fields in the $x^{\ha}$-directions are as
in (\ref{3.6}). The spectrum for directions other than the
$x^{\ha}$-directions is just standard. For the
$x^{\ha}$-directions, and using
the above results, we have $\e=0$ and therefore the
fermionic worldsheet excitations are also standard. For the bosonic
fields the only relevant difference is the corresponding zero
modes which become commuting variables and no longer satisfy the
commutation relation (\ref{3.20}).\footnote{Note that taking the
  limit $\g=\g'$ the commutation relation (\ref{3.20}) becomes ill
  defined. However, in this limit the spectrum of the Landau levels
  becomes continuum as may be seen from (\ref{3.22}). This is the
  expected result because the zero modes are now continuum
  variables. We thank Gary Gibbons for bringing this point to our
  attention.} 
Also, the operators $a_{\e}$ and $a_{\e}^{\dagger}$ disappear and are
substituted by a linear term in $\t-if\s$ \cite{Abou..}. The
resulting spectrum is exactly the same as in the case for vanishing
worldvolume gauge field (up to a normalisation factor in the momentum
modes that is correctly reproduced for the low-lying excitations if
one considers the Born-Infeld action \cite{HashTayl}).

A crucial difference between the open superstrings obeying the
boundary conditions (\ref{3.39}) and the boundary conditions corresponding
to a D-$p$-brane with vanishing worldvolume gauge field is the vacuum
amplitude for two parallel of such branes. In the bound state case
there is a overall factor $(1+f^2)$ multiplying the corresponding
result for two D-$p$-branes. This factor arises in the momentum
integration as it was explained in \cite{Abou..}. In the effective
field theory this factor also appears as it is expected from the
agreement of both calculations \cite{Polc}. Consider first the
attractive term in the amplitude that is due to the graviton,
dilaton and antisymmetric tensor field exchanges. The coupling of a
D-brane to this fields will
have an extra $\sqrt{1+f^2}$ factor due to the Born-Infeld character
of the D-brane action \cite{Green}, giving the correct result in the
corresponding
one-loop calculation. The repulsive RR exchange will now have two
contributions as the D-branes carry the charge of the ${\cal A}_{p+1}$
and ${\cal A}_{p-1}$ form field potentials. To be more precise, the
coupling of a D-$p$-brane to these fields will be given by
\eqn
\arr{l}  
\mu_p\int d^{p+1}\s
\left( \hat{{\cal A}}_{p+1}+F\w\hat{{\cal A}}_{p-1}\right) =\\\\
\mu_p\int d^{p+1}\s\hat{{\cal A}}_{p+1}+
\mu_p f \left( 2\pi R_{p-1}s\right)
\left( 2\pi R_pp\right)\int d^{p-1}\s\hat{{\cal A}}_{p-1}\ ,
\earr
\label{3.40}
\eeqn
where the hat denotes the pullback to the D-brane's worldvolume,
$\mu_p$ is the D-$p$-brane quantum of charge and this D-brane is
wrapped around the $x^{p-1}$- and $x^p$-directions with winding
numbers $s$ and $p$, respectively. Both the RR exchanges will give the
desired $1+f^2$ factor. We remark that 
since just integer quanta of the ${\cal A}_{p-1}$ field can be
exchanged there has to be a quantisation condition on the worldvolume gauge
field. In other words, the usual flux quantisation condition has to be
satisfied. This gives the relation
\eqn
q\mu_{p-2}=\mu_p f ( 2\pi R_{p-1}s)( 2\pi R_pp)\ ,
\label{3.41}
\eeqn
where $q$ is an integer. Substituting for the known values of $\mu_p$
and $\mu_{p-2}$ we find
\eqn
f=\frac{1}{s}\frac{q}{p}\frac{\a'}{R_{p-1}R_p}\ .
\label{3.42}
\eeqn
This is exactly the quantisation condition (\ref{2.8}) for the case
$s=1$ obtained geometrically be performing $T$-duality at an angle.
If $s\ne 1$ the corresponding T-dual D-brane system is given by $s$
parallel branes wrapped on a $(q,p)$-cycle on this 2-torus.

\section{D-brane entropy counting}
\news

In this section we will study the D-brane system describing the
extreme and near-extreme black hole presented in section 2. This
discussion parallels that of \cite{CallMald, Mald} with some new
ingredients. We start by performing the D-brane entropy counting in
the extreme case. Our D-brane system consisted of $N_1$ and $N_2$
intersecting D-$4$-branes with the $U(1)$ center of the corresponding
gauge fields turned on. It was represented as
\eqn
\arr{rl}
4_1:&X^1,X^2,X^4,X^5,\\
    &F_{45}^{(1)}=\tan{\z}\ ,\\
4_2:&X^1,X^3,X^4,X^5,\\
    &F_{45}^{(2)}=-\cot{\z}\ ,
\earr
\label{4.1}
\eeqn
where the branes carry $N$ units of left-moving momentum in the
$x^1$-direction. The D-4-branes are also wrapped on the
$x^5$-direction with windings $p_1$ and $p_2$.
The excitations of this D-brane system are described
by three different sectors. The first two correspond to open strings
with both ends on the same group of D-branes and the third to open
strings with each end on a different group of D-branes. The former
cases were explained at the end of the previous section and the
spectrum is similar to the case with vanishing worldvolume gauge
field. The maximum number of massless excitation is $8N_i$ (and
not $8N_i^2$ because some excitations will give a mass term to others)
which is far to few to explain the entropy formula (\ref{2.16})
\cite{Mald}.  This excitations represent subleading contributions to
the entropy formula. We expect these string states
to condense ensuring supersymmetry of the excited D-brane system. In
the gauge theory description this fact translates into the conditions
for the vanishing of the D-terms of the theory and of the mass terms
for the hypermultiplet associated with strings ending on different
branes\cite{Mald}. The relevant excitations come from the sector of
open strings with the ends on different branes \cite{CallMald}.
Open strings attached to both type of branes will
have the following boundary conditions
\eqn
\arr{ll}
X^0,\ X^1\ :& NN,\\
X^2,\ X^3\ :& ND,\ DN,\\
X^4,\ X^5\ :& FF,\\
X^6,...,X^9\ :& DD,\\
\earr
\label{4.2}
\eeqn
where the FF boundary conditions represent open strings satisfying the
boundary conditions (\ref{3.5}) and (\ref{3.6}) with the field
strength given by (\ref{3.4}) and (\ref{3.4a}). The only
non-standard results come from these $x^4$- and $x^5$-directions. The
parameters in the previous section read now in terms of our D-brane
system as
\eqn
\arr{c}
\g =\z\ ,\ \ \ \g' =\z-\frac{\pi}{2}\ ,\ \ \ \e=\frac{1}{2}\ ,\\
n_L=p_2q_1+p_1q_2\ .
\earr
\label{4.3}
\eeqn

Consider first the Neveu-Schwarz sector of the theory. Since
$\e =1/2$, the worldsheet bosons become half-integer moded and
the worldsheet fermions integer moded. The zero moded operators
$\Psi_0$, $\Psi^{\dagger}_0$ take ground states into ground
states. Thus, the mode expansion is just as in the ND case. The zero
energy is then seen to be zero as it is the case of intersecting
D-branes with four ND directions
\eqn
E=(6-2)\left( -\frac{1}{24}-\frac{1}{48}\right)+
4\left( \frac{1}{48}+\frac{1}{24}\right) =0\ ,
\label{4.4}
\eeqn
where the first term is the contribution of the
$x^0,x^1,x^6,...,x^9$-directions and the second term the contribution
of the DN, ND and FF directions.\footnote{The zero energy in the FF
  directions is shifted by $\e/2=1/4$ with respect to the
  usual result for two NN or DD directions as it was explained in the
  previous section.}
There is however a crucial difference between
the ND and FF directions since the latter admits a degenerated zero
mode in the bosonic expansion. The vacuum state for the NS sector
forms a representation of the $4$-dimensional algebra
\eqn
\left\{ \Psi_0^I,\Psi_0^J\right\} =\d ^{IJ}\ ,\ \ \ I,J=2,...,5,
\label{4.5}
\eeqn
and it has a further degeneracy due to the bosonic zero mode. A
convenient basis for the vacuum states is
\eqn
|s_1,s_2,y_m\rangle\ ,
\label{4.6}
\eeqn
where $s_i=\pm\frac{1}{2}$ and $y_m$ comes
from the Landau
degeneracy as explained in subsection 3.3. This vacuum transforms as a
spinor under the internal space group $SO(4)_I$ giving therefore
bosonic states. The GSO projections leaves half of the states out. We
end up with $2n_L$ states, and since there are $2N_1N_2$ of these
strings (the factor of $2$ is because the strings are oriented) we have
$4N_1N_2n_L$ massless bosonic particles.

Next, consider the Ramond sector. The zero energy is zero as it is
usual even when we have the FF directions. Since $\e=1/2$, the
worldsheet fermions become half-integer moded as it is the case for ND
directions. This vacuum forms a representation of the 6-dimensional
Dirac algebra
\eqn
\left\{ \Psi_0^{\mu},\Psi_0^{\nu}\right\} =\eta ^{\mu\nu}\ ,
\ \ \ \mu,\nu=0,1,6,...,9,
\label{4.7}
\eeqn
with the Landau degeneracy. A basis for the vacuum states is
\eqn
|s_1,s_2,s_3,y_m\rangle\ ,
\label{4.8}
\eeqn
where $s_i=\pm \frac{1}{2}$ and $y_m$ is defined as in
(\ref{4.6}). This vacuum transforms as a spinor under the corresponding
$SO(1,5)$ group. The GSO projection plus the fact that physical states
are annihilated by the zero mode of the supersymmetry generator
leaves us with $2n_L$ states and therefore we have $4N_1N_2n_L$
massless fermionic excitations.

The argument now is exactly as presented in \cite{CallMald}. We have a
gas of strings with $4N_1N_2n_L$ fermionic and bosonic species
carrying all together the left-moving momentum
$P=\frac{N}{R_1}$. This gives exactly the entropy formula
(\ref{2.16}). 

The previous counting argument is valid for $N\gg n_LN_1N_2$
\cite{MaldSuss}. Outside this regime, we would obtain
an entropy much smaller than the Bekenstein-Hawking entropy. The point
is that the  configurations of branes
winding around in the $x^1$-direction become the relevant ones to
explain the classical geometry limit.
Consider the case where $N_1$ and $N_2$ are co-prime (this assumption
may be easily relaxed \cite{MaldSuss}). For the system corresponding
to two intersecting D-4-branes with winding numbers $N_1$ and $N_2$ in
the $x^1$-direction we will have a gas of 
$4n_L$ bosonic and fermionic species. The momentum carried by these
states is however quantised in units of $(N_1N_2R_1)^{-1}$ giving the
correct entropy formula for $NN_1N_2\gg n_L$.

\subsection{Near-extremal black hole}

In the extremal case we can extrapolate the entropy calculation from
the weakly coupled D-brane phase to the strongly coupled classical
black hole geometry phase because of supersymmetry. In other words,
the BPS nature of our configuration allow us to extrapolate between
small and strong coupling regimes while leaving the number
(degeneracy) of our BPS states unchanged. For the non-extreme solution
case this argument is no longer valid. While the
small coupling calculation is still reliable in the non-BPS case, we
can not extrapolate it to the strong coupling region. However, it is
seen that in the near-extreme case the D-brane picture reproduces the
correct semi-classical thermodynamical analysis \cite{CallMald}. As
explained in \cite{MaldSuss}, the excitations of our D-brane system
above the BPS state that correctly reproduce the near-extreme
behaviour correspond to multiply wrapped branes. 
The corresponding D-brane system is defined by keeping the total
momentum and D-brane charges fixed, while introducing some
right-moving momentum $\d N_R (=\d N_L)$, and anti-D-$4$-branes both
with magnetic condensates as in (\ref{4.1}) and with windings along
the $x^1$-direction $\d N_{\bar{1}} (=\d N_1)$ and $\d N_{\bar{2}}
(=\d N_2)$.

The calculation of the variation of the entropy due to the change in
momentum is similar to the one presented in
\cite{CallMald,MaldSuss}. We have $4n_L$ bosonic and fermionic species
with momentum quantised in units
of $(N_1N_2R_1)^{-1}$. The change in the left-moving
entropy is $\Delta S=2\pi\sqrt{N_1N_2n_L}
(\sqrt{N+\d N_R}-\sqrt{N})$, while the change in the right-moving
entropy is $\Delta S=2\pi\sqrt{N_1N_2n_L}\sqrt{\d N_R}$ which
dominates. The resulting change in the entropy agrees with the
semi-classical value in (\ref{2.28}).

We proceed by calculating the contribution to the entropy due to the
anti-D-4-branes. In order to do that we shall assume that in the
near-extreme regime we may perform duality transformations leaving the
degeneracy of states unchanged \cite{CallMald}. Consider the case when
we have a D-4-brane and an anti-D-4-brane along the
$x^1,x^2,x^4,x^5$-directions with a magnetic condensate as in
(\ref{4.1}) and wrapped on the $x^1$-direction with winding numbers
$N_1+\d N_{\bar{1}}$ and $\d N_{\bar{1}}$, respectively. Next, perform
the following duality transformations
\eqn
T_1\ S\ T_4\ T_5\ T_2\ (Up)\ ,
\label{4.9}
\eeqn
where $T_i$ means a T-duality transformation along the
$x^i$-direction, $S$ the S-duality transformation and $(Up)$ the uplift of
the configuration to 11-dimensions. To follow this duality orbit one
has to realize that S-duality acts as electromagnetic duality on the
worldvolume gauge field of the D-3-branes \cite{Tsey,GreenGutp1}. The
duality transformation (\ref{4.9}) interchanges the $N$ quanta of
momentum in the $x^1$-direction with a M-5-brane parallel to the
$x^1,x^4,x^5,x^2,x^{11}$-directions with winding number $N$ along the
$x^{11}$-direction. The D-4-brane with winding $N_2$ in the
$x^1$-direction is transformed into a membrane singly wrapped on the
$x^3$-direction and wrapping diagonally the $x^2x^{11}$ 2-torus on the
$(p_2,-q_2)N_2$-cycle \cite{RussoTsey}. This membrane makes an angle
$-\z$ with the $x^{11}$-direction in this squared 2-torus where
\eqn
\cot{\z}=\frac{q_2}{p_2}\frac{R_{11}}{R_2}\ .
\label{4.10}
\eeqn
The M-5-brane intersects the membrane along the string defined by this
direction. The D-4-brane with winding $N_1+\d
N_{\bar{1}}$ (anti-D-4-brane with winding $\d N_{\bar{1}}$) is
transformed under the
duality operation (\ref{4.9}) into left(right)-moving momentum modes
propagating along the common string direction. There will be
$p_1(N_1+\d N_{\bar{1}})$ left-moving quanta of momentum along the
$x^{11}$-direction and $q_1(N_1+\d N_{\bar{1}})$ left-moving quanta of
momentum along the $x^2$-direction. The condition that these momentum modes
propagate along the common string is
\eqn
\tan{\z}=\frac{q_1}{p_1}\frac{R_{11}}{R_2}\ .
\label{4.11}
\eeqn
The total left-moving momentum along this direction is then seen to be
\eqn
P_L=\frac{(N_1+\d N_{\bar{1}})n_L}{RNN_2}\ ,\ \ \ 
R=\sqrt{(q_2R_{11})^2+(p_2R_2)^2}\ .
\label{4.12}
\eeqn
We now note that the momentum modes along the intersection string are
described by a superconformal field theory with central charge $c=6$
\cite{KlebTsey,BalaLars}. In fact, reducing our M-theory
configuration along the $x^4$-direction we obtain a D-4-brane
intersecting a D-2-brane over the string direction described
above. The open strings carrying the momentum along this string
direction are described by a theory with the required central
charge. Thus, from equation (\ref{4.12}) we conclude that the left
sector of the theory is at level $NN_2(N_1+\d N_{\bar{1}})n_L$ giving
a change in the entropy $\Delta S=2\pi\sqrt{NN_2n_L}
(\sqrt{N_1+\d N_{\bar{1}}}-\sqrt{N_1})$. Similarly, the change in the entropy
arising from the right-moving sector is $\Delta
S=2\pi\sqrt{NN_2n_L}\sqrt{\d N_{\bar{1}}}$ which dominates. The resulting
change in the entropy again agrees with the semi-classical result
(\ref{2.28}). The near-extreme entropy due to the other
anti-D-4-brane may be calculated in a similar way.

The calculation of the Hawking temperature goes through as in
\cite{CallMald}. The result again agrees with (\ref{2.28}) and is
\eqn
T_H =2T_R = \frac{4}{2\pi R_1}\sqrt{\frac{\d N_R}{N_1N_2n_L}}\ ,
\label{4.13}
\eeqn
where $T_R$ is the right-moving temperature.

\section{Conclusion}

In this paper we have studied a black hole described by a
configuration of intersecting D-4-branes with non-vanishing
worldvolume gauge fields. We have shown the agreement between the
semi-classical and D-brane calculations of the entropy and Hawking
temperature in the extreme and near-extreme cases. As a new
ingredient, we found that the Landau degeneracy of open string states
describing the excitations of the D-brane system enters in a
fundamental way in the explanation of the microscopic structure of this
black hole. It is amusing that an old result of quantum mechanics that
has been brought into string theory in \cite{Abou..} now has its place 
in the fundamental description of black holes, providing an impressive
new matching with the corresponding semi-classical results.

The results presented in this paper should also arise in the
worldvolume description of the low-lying states of our D-brane
system. The corresponding theory should be dual to an 8-dimensional
theory with two vector gauge fields $A_{\a}$ ($\a=0,...,7$) and two
scalar fields $\phi_m$ ($m=1,2$) in the adjoint representation of
$U(N_1p_1)$ and $U(N_2p_2)$, plus a hypermultiplet transforming in the
fundamental representation of $U(N_1p_1)\times U(\overline{N_2p_2})$.
Naively one should expect from the D-brane intersection rules that
such a supersymmetric theory does not exist. However, the coupling to
the non-vanishing background gauge fields should render the theory
supersymmetric. 

As another application of our work there is the possibility of
considering the dual configuration of $n$ D-branes intersecting at $SU(2)$
angles \cite{Berk..,Brec..2,Bala..}. The entropy of such configuration
of D-5-branes with a self-dual field strength on a compact $T^4$ also
has its origin in the
Landau degeneracy of open strings describing the excitations of
this system. Further, in this case it is easy to write down the
worldvolume field theory  governing the dynamics of the
D-brane system. It is just the super Yang Mills theory with twisted
boundary conditions on $T^4$. The Landau levels arise in this
description as torons \cite{Baal,HashTayl}, and the correct matching
with the entropy formula arises from non-trivial results on
$\Theta$-functions on $T^4$ \cite{CostaPerry}. 

Finally, let us note that the intersection of other D-branes with
worldvolume gauge fields may be studied 
by using the results of section 3. An example, are the
configurations found in \cite{Costa1}. The corresponding worldvolume
theory should also be expressed in terms of bundles with twisted
boundary conditions.

\section*{Acknowledgements}
We would like to thank M. Gutperle and G. Gibbons for helpful
comments. One of us (M.S.C.) acknowledges the financial support of
JNICT (Portugal) under programme PRAXIS XXI.

\end{document}